\documentclass[conf]{new-aiaa}
\usepackage[utf8]{inputenc}

\usepackage{graphicx}
\usepackage{amsmath}
\usepackage[version=4]{mhchem}
\usepackage{siunitx}
\usepackage{longtable,tabularx}
\usepackage{physics}
\usepackage{xcolor}
\usepackage{csvsimple}
\usepackage{subcaption}
\usepackage{threeparttable}

\title{Generation of High-Order Coarse Quad Meshes\\
on CAD Models via Integer Linear Programming}

\author{
    Mattéo Couplet\footnote{
       F.R.S.-FNRS Research Fellow, \href{mailto:matteo.couplet@uclouvain.be}{matteo.couplet@uclouvain.be}.},
    Maxence Reberol\footnote{
        Postdoctoral Researcher, \href{mailto:maxence.reberol@uclouvain.be}{maxence.reberol@uclouvain.be}.} and 
    Jean-François Remacle\footnote{
        Professor, Louvain School of Engineering, \href{mailto:jean-francois.remacle@uclouvain.be}{jean-francois.remacle@uclouvain.be}.
    }
}
\affil{Université catholique de Louvain, Louvain-la-Neuve, Belgium}

\begin{document}

\maketitle

\begin{abstract}
    We propose an end-to-end pipeline to robustly generate
    high-quality, high-order and coarse quadrilateral meshes on CAD models.
    This kind of mesh enables the use of high-order analysis techniques
    such as high-order finite element methods or isogeometric analysis.
    An initial unstructured mesh is generated;
    this mesh contains a low number of irregular vertices
    but these are not necessarily aligned,
    causing a very dense quad layout.
    A T-mesh is built on the mesh which allows to modify
    its topology by assigning new integer lengths to the T-mesh arcs.
    The task of simplifying the quad layout can be
    formulated as an Integer Linear Program
    which is solved efficiently using an adequate solver.
    Finally, a high-order quad mesh is extracted from the optimized topology.
    Validation on several CAD models
    shows that our approach is fast, robust,
    strictly respects the CAD features,
    and achieves interesting results in terms of coarseness and quality.
\end{abstract}

\section{Introduction}
\vspace{1em}
% Critical because reviewers and readers
% (i) form their opinion from it
% (ii) get back to it regularly
% Goal: give a more extensive sense of what paper is about

%\subsection{Context and motivation}
% Context of the problem addressed
% Overview of relevant work in the area
% Definition of the problem, motivation for addressing it

% Why do we want quad meshes? Why HO quad meshes in particular? Applications?
Computer Aided Design (CAD) systems are used extensively for
industrial design in many domains, including automotive,
shipbuilding, aerospace industries, industrial and architectural
design, prosthetics, and many more.
Engineering designs are encapsulated in CAD models, which, up to
manufacturing precision, \emph{exactly represent their geometry}.
While the engineering analysis process begins with such CAD models,
the predominant method of analysis (the finite element method)
requires an alternative, discrete, representation of the geometry:
a finite element mesh. In a mesh, the CAD model is usually
subdivided into a large collection of simple geometrical
shapes. Mesh generation can thus be seen as the transmission belt between
CAD and analysis. Notice that, even though the mesh is not geometrically as precise as the CAD,
mesh data are usually way larger than CAD data.

The surface of a CAD model is typically defined by piecewise-smooth CAD
patches. CAD models are filled with various idiosyncrasies:
periodic surfaces, trimmed patches, exotic or singular parametrizations. 
The idea of this paper is to provide a representation of the CAD that is
light, universal and precise. For that, we build a high-order quad layout of the whole
CAD model. A quad layout is the partitioning of an object's surface
into simple networks of conforming quadrilateral patches. A high-order
quad layout is a quad layout where individual patches are endowed with
a geometry. Here,
individual patches are both i) truly smooth and ii) share the same
topology (four sides, no internal boundaries). The geometry of individual high order
quad patches can be described either using high order finite elements or can inherit from
the CAD representation (e.g. surface B-splines). The quad patches can be
aligned with some of the feature edges of the CAD model but can also ignore
some.

Such a coarse high-order quad layout can be used as a CAD model but also as a mesh
in a high-order finite element context ($p$-FEM), or in numerical techniques
with NURBS-based representation of the geometry, such as isogeometric analysis
\cite{hughes2005} and NEFEM~\cite{sevilla2008}. It can also be used a
lightweight datastructure for forming an inline block-structured quad mesh.
Block-structured meshes  are of great interest for geometric multigrid or fast
to design computation kernels.

%High-order quadrilaterals could be used to accurately capture the
%underlying geometry of the patches and thus serve as a simplified CAD
%model that is easily understandable by finite element
%practitioners. The word simplified should not be taken in the sense
%of a simplification of the geometry: 

%using high order quads allows to
%produce a high fidelity CAD model with 

%%For several decades, the engineering analysis community has been interested
%in generating quadrilateral meshes for CAD models,
%as they form a good geometrical support
%for finite element and finite volume methods.
%Even more ideal are high-order quadrilateral meshes, i.e.,
%meshes consisting of fewer but coarser high-order quadrilaterals.
%We see three useful applications to such meshes.

%% FIXME -- HERE WE COMPUTE A  LAYOUT --
% Standard technique to build a HO mesh and why it's not ideal.
%A classical technique to build a high-order mesh is first
%to generate a coarse linear mesh, then deform it to match the CAD geometry.
%However, due to the many small features that common CAD models possess,
%a coarse quadrilateral mesh directly built on them will present
%highly distorted elements or very irregular vertices,
%severely impacting the resulting high-order mesh.

% Our approach, in short
In this paper we propose a practical approach to the
generation of high-quality high-order quad layouts,
where the emphasis is on robustness with respect to complex CAD models
commonly occurring in engineering analysis.
Our approach consists in first generating
an unstructured linear and fine quadrilateral mesh,
then modifying its topology until it is possible to extract a coarse
quad layout.

%\subsection{Problem statement}

% What is the input?
Three-dimensional CAD models are represented on a computer
using a \emph{Boundary Representation} (BRep): a volume is
bounded by a set of faces, a face is bounded by a series of
curves and a curve is bounded by two end points.
The BRep is a discrete object: it is a graph that contains model
entities together with all their topological adjacencies.
Then a geometry is associated to each model entity.
Here, we aim at handling \emph{real life} CAD models occurring
in engineering analysis, which serve as a support
for finite element or finite volume methods.
Such models may be arbirtary complex, because of their size (large number of
model entities) and/or because of the existence of
sharp angles and features at very different scales. This complexity
makes the robust treatment of CAD models particularly challenging.

In this work, we make the following assumption:
the model is assumed to be a 2-manifold. Each model curve
must be adjacent to at most two model faces. This
limitation is due to implementation difficulties and could be overriden
in a future work.
% What do we want to produce? What are the quality criteria?
At first let us define specifications for the high-order quad layout.
\vspace{.3cm}
\begin{enumerate}
    \item The layout should be \emph{conforming}, i.e., free of
      T-junctions. This is a necessary condition for most applications
      (multiblock meshing, isogeometric analysis). 
    \item The quadrilaterals of the layout should strictly \emph{conform to the CAD},
    i.e., the quad corners should exactly lie on the underlying model surfaces
    and quad edges should strictly be aligned with a set of user-flagged model curves.
    \item The mesh should be \emph{valid} and, to a certain extent,
    \emph{geometry-consistent}; the exact meaning of these terms will be detailed later.
    \item The layout should be \emph{coarse};
    hence our primary concern is to minimize the number of quads in
    the layout.
\end{enumerate}
\vspace{.3cm}

The idea here is to start from a quad layout that does not fulfill all
specifications and then to transform it to obtain compliance with the specifications.
In our approach, the starting point is a valid quad mesh: a quad
layout can be extracted from a conforming quad mesh and this layout verifies 
all expected specifications except the last one: a quad mesh does not
usually produce a coarse layout (see \S\ref{sec:quadqs} and Figure \ref{fig:input_layout}).
This is mainly caused by non-alignment of the irregular vertices.

\subsection{Related work}

The generation of coarse quadrilateral layouts, also called base-complex in the
literature, is a well-studied problem. For planar models and open surfaces,
starting from a fine quad mesh, a simple approach is to iteratively collapse
the chords (dual quad loops) which separate the irregular vertices that should
be aligned. Unfortunately, this does not generalize to closed surfaces because
the chords tend to wind up around the whole model, and collapsing them would
simply destroy the mesh.

Bommes et al.~\cite{bommes2011} automatically detect helical configurations,
i.e., quad loops forming helices winding around the model, and correct them by
modifying the mesh directly using elementary operations (such as collapsing and
shifting edges) that preserve the quad-nature of the mesh. But this approach is
not sufficient to deal with many other configurations, such as self-intersecting quad
loops.

Tarini et al.~\cite{tarini2011} extract the quad layout
from the input mesh and simplify it by iteratively
changing connections between irregular vertices in a greedy fashion,
in order to minimize concurrently the separatrices' length and drift
 (how much they deviate from the underlying cross field).
Once the new separatrices are optimized,
the mesh is re-parametrized to align it with the new layout.

Razafindrazaka et al.~\cite{rafa2015} optimize the network of separatrices
connecting irregular vertices by formulating it as a graph matching.  The
resulting network is optimal with respect to a user-specified balance between
coarseness and geometric feature alignment.

An alternative approach is to directly produce the coarse quadrilateral layout
from a seamless parametrization (consistent global integration of a
cross-field). Lyon et al.~\cite{lyon2021} first build a T-mesh (non-conforming
quad layout with T-junctions) from a seamless global $uv$-parametrization.
The integer lengths of the T-edges are then optimized to achieve a conforming
coarse layout. The benefit of this approach is that it encodes an infinite
number of layout connections in a finite structure, and allows to flexibly
impose constraints.

\vspace{.3cm}

These papers have in common that they focus on computer graphics applications,
where CAD consistency is generally not a concern and where the number of feature
curves is limited. Our work aims at achieving similar results but with the
additional constraint that the quad layout must comply with a complex CAD
model. 

\subsection{Overview}

For the construction of the coarse layout, we build on the recent quantization
simplification of Lyon et al.~\cite{lyon2021} as it provides the most
flexibility to impose the constraints related to CAD compliance. Besides
additional CAD constraints, the main difference with their work is that we use
only quadrilateral meshes: neither a global seamless parametrization nor
a cross field is needed here.  In a CAD context, we believe that starting from an unstructured
quad mesh is more robust as it is much easier to generate one compared to a valid
global seamless parametrization.

\begin{figure}[htbp]
    \centering
    \begin{subfigure}{0.33\linewidth}
        \centering\includegraphics[width=\linewidth]{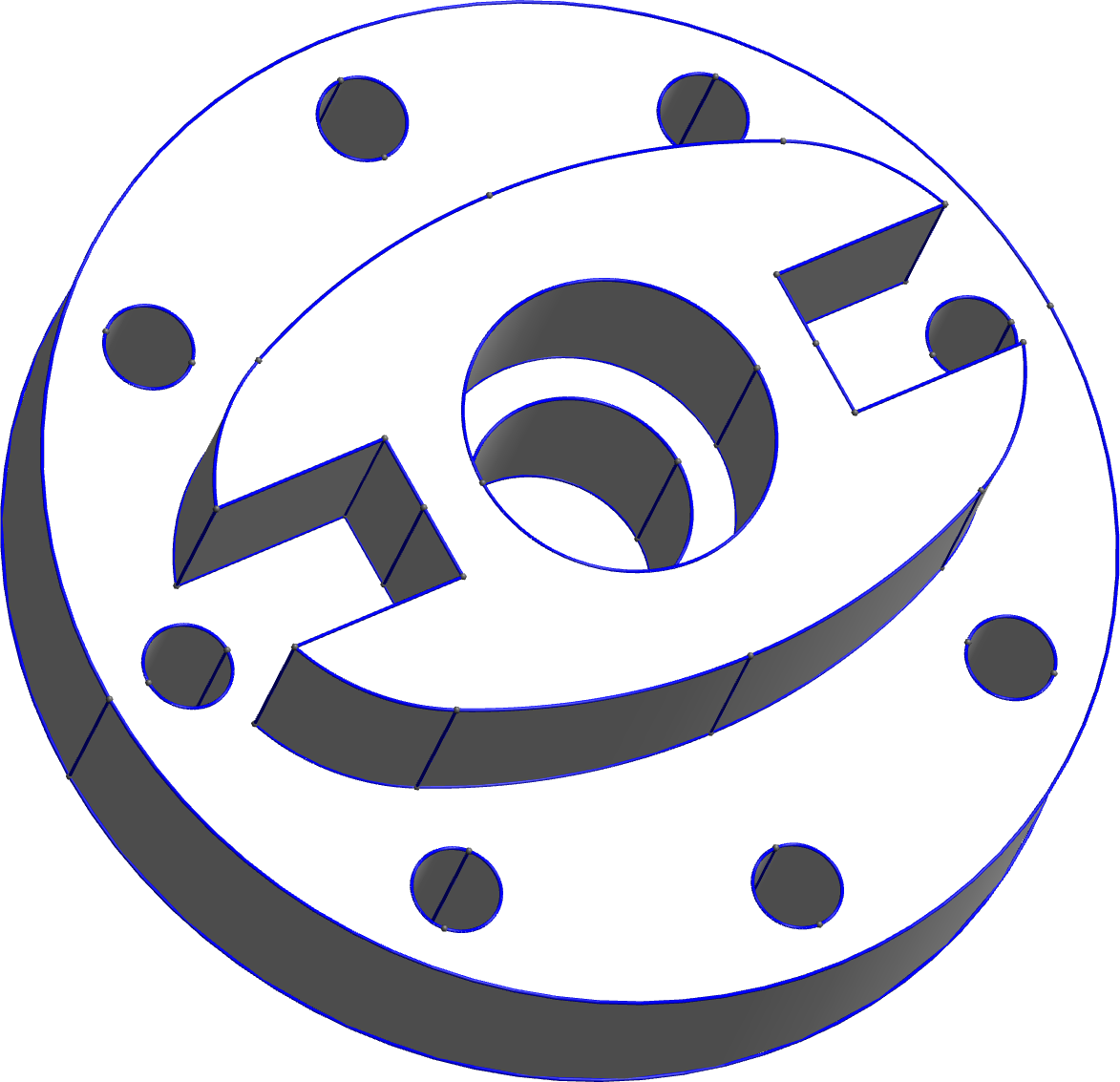}
        \caption{CAD model}
    \end{subfigure}
    \begin{subfigure}{0.33\linewidth}
        \centering\includegraphics[width=\linewidth]{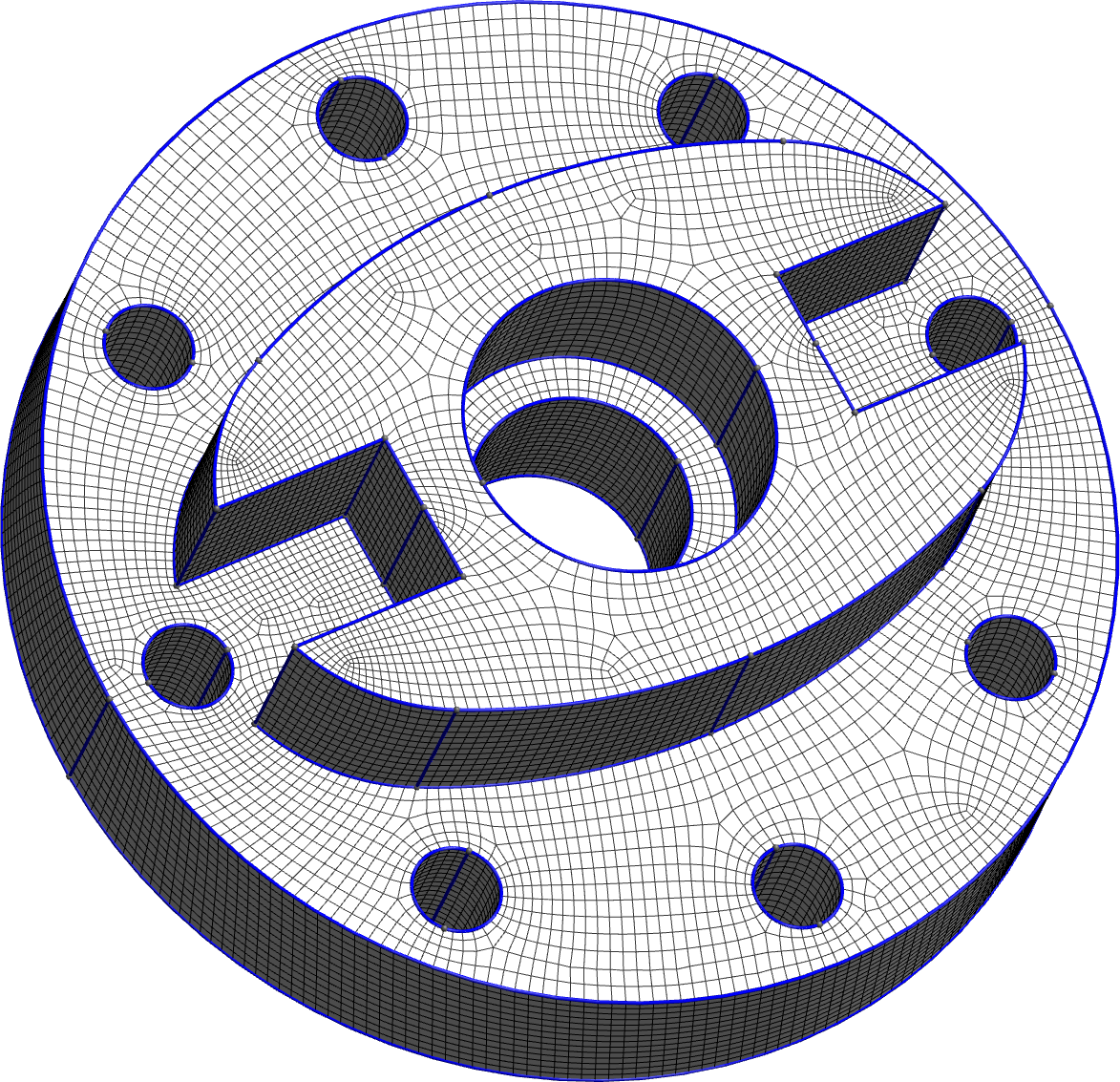}
        \caption{Unstructured quad mesh}
    \end{subfigure}
    \begin{subfigure}{0.33\linewidth}
        \centering\includegraphics[width=\linewidth]{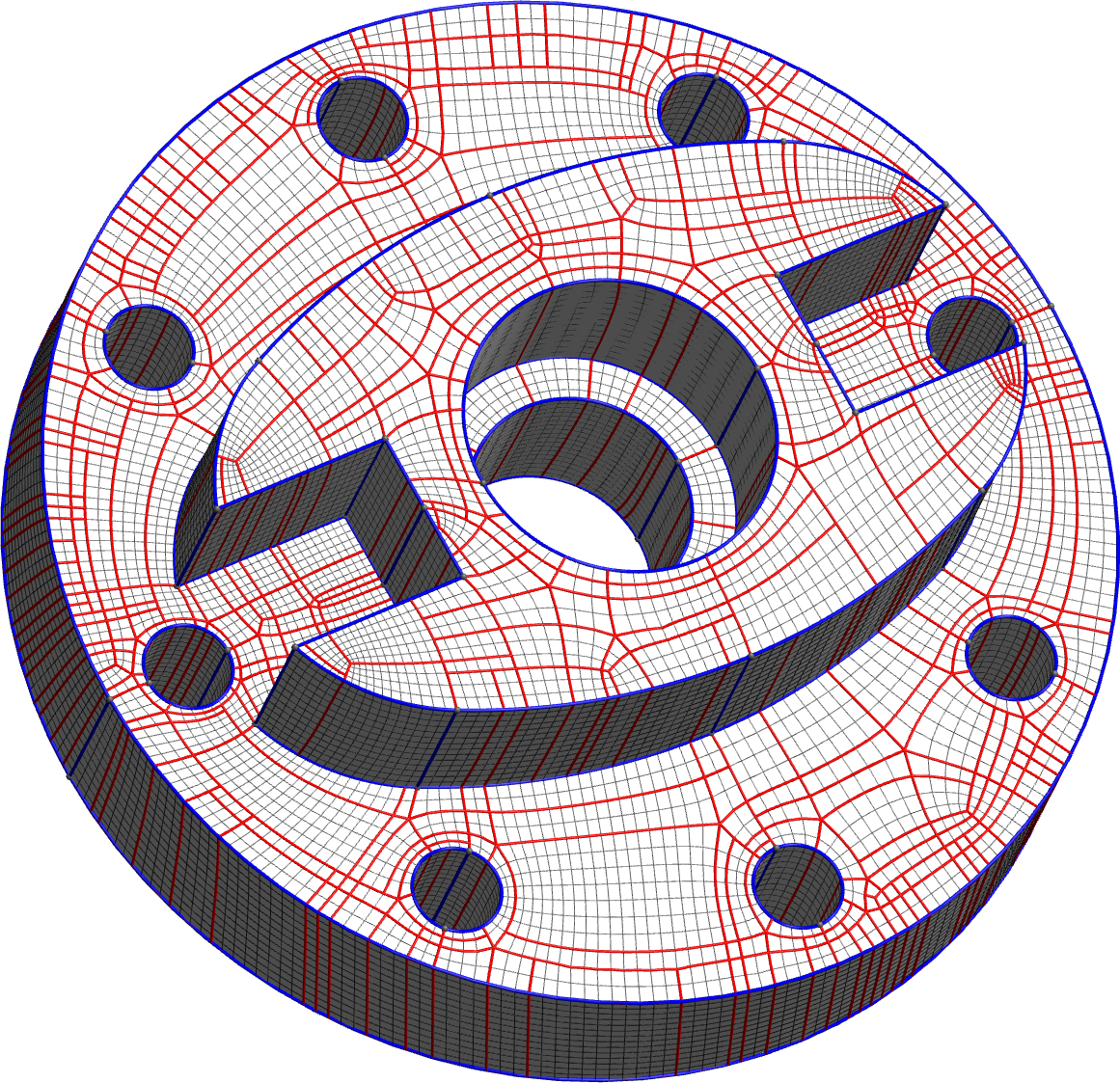}
        \caption{T-mesh}
    \end{subfigure} \\
    \begin{subfigure}{0.33\linewidth}
        \centering\includegraphics[width=\linewidth]{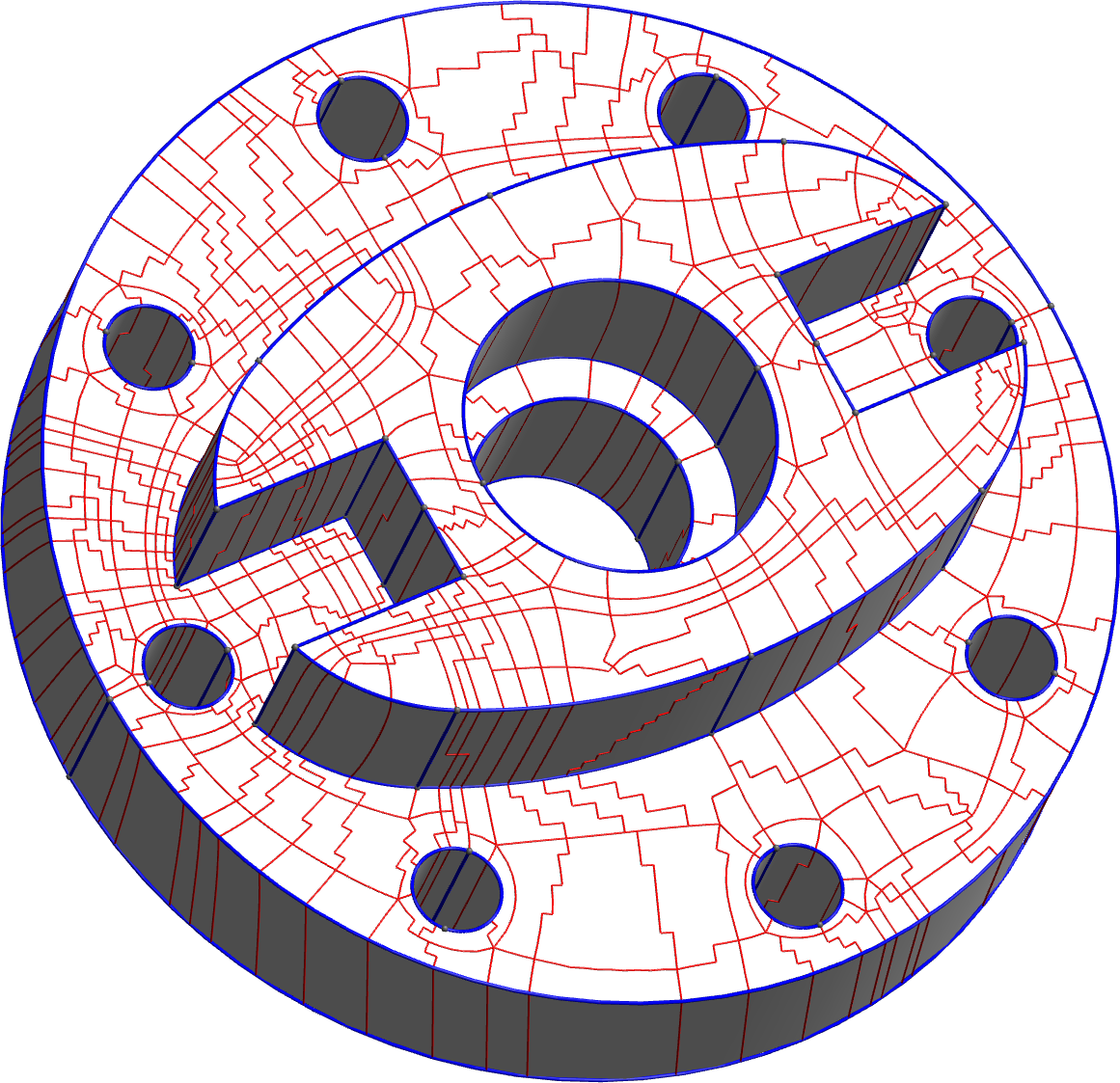}
        \caption{Coarse layout}
    \end{subfigure}
    \begin{subfigure}{0.33\linewidth}
        \centering\includegraphics[width=\linewidth]{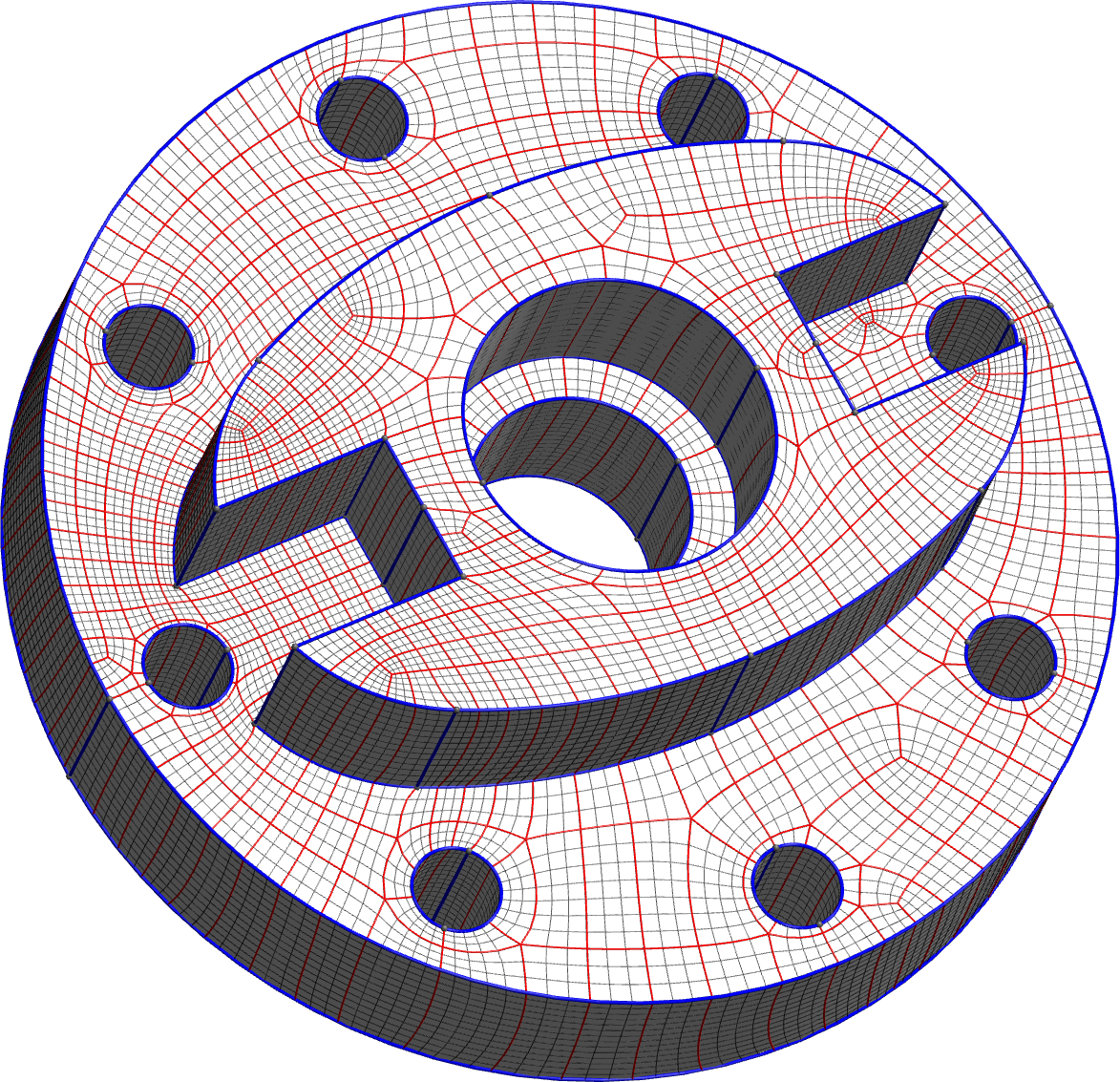}
        \caption{Block-structured mesh}
    \end{subfigure}
    \begin{subfigure}{0.33\linewidth}
        \centering\includegraphics[width=\linewidth]{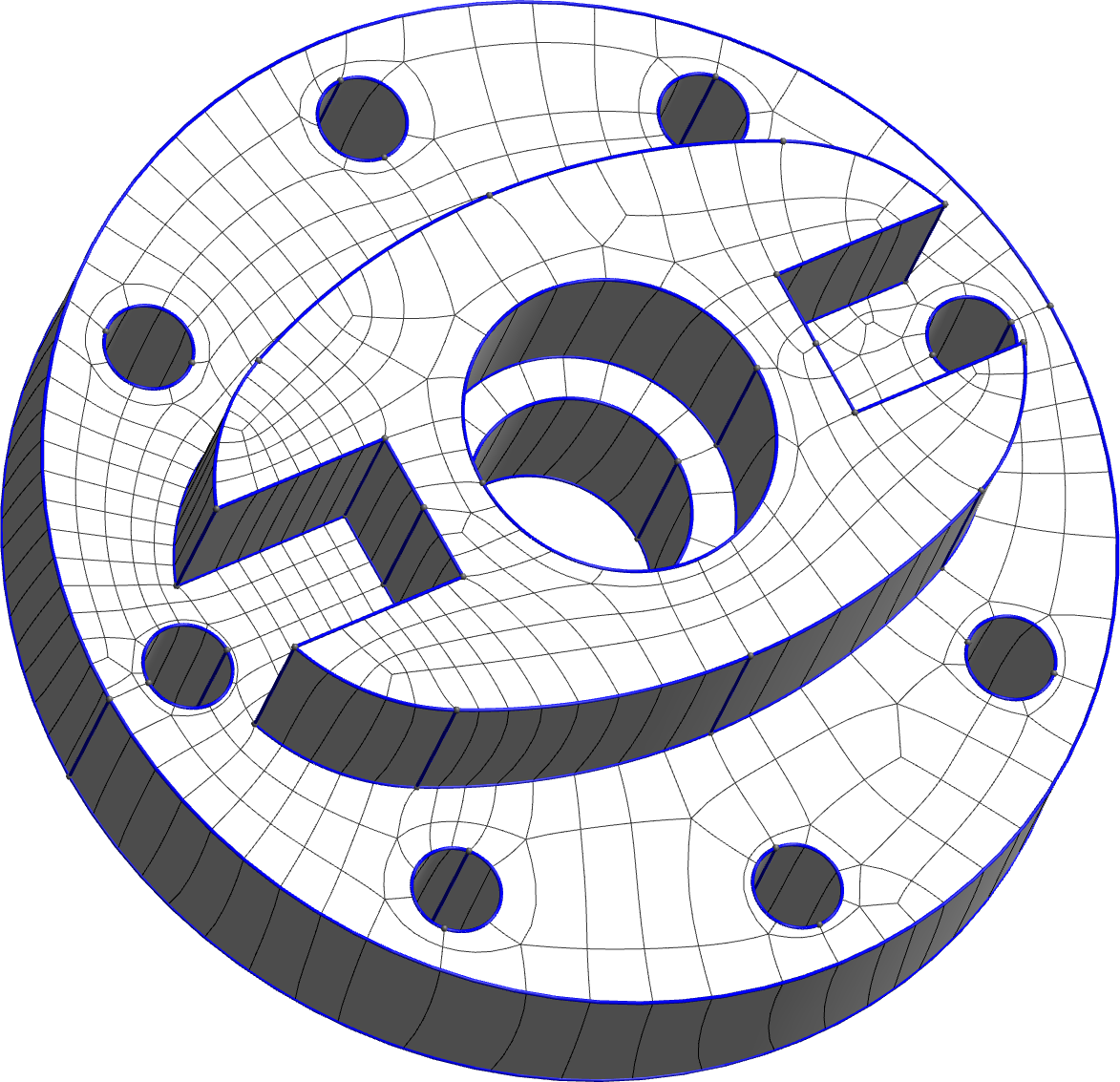}
        \caption{High-order mesh}
    \end{subfigure}
    \caption{Overview of our high-order quadrilateral meshing pipeline.
    The illustrated CAD model is M1 from the MAMBO dataset~\cite{mambo}.}
    \label{fig:overview}
\end{figure}

Our complete method for building a high-order quadrilateral layout from a CAD model
works as follows, and is illustrated in~\autoref{fig:overview}:
\begin{enumerate}
    \item First, an \emph{unstructured} quad mesh is built on the model.
    This step relies on our previous work~\cite{reberol2021};
    see Section~\ref{sec:quadqs}.
    \item Then, the topology of the mesh is modified 
    to obtain a coarse layout,
    using our adaptation of the T-mesh approach of~\cite{lyon2021};
    see Section~\ref{sec:layout}.
    \item The geometry of the coarse layout is optimized by quantizing the
        patches and smoothing the resulting fine mesh;
        see Section~\ref{sec:bsquad}.
    \item High-order quadrilaterals are eventually extracted from
        the block-structured fine mesh and the CAD geometry; see Section~\ref{ssec:extraction}.
\end{enumerate}

%\subsection{Related work}
% What have others done to address the stated problem?
% What does this paper bring?

% \section{Definitions}
% What is a mesh? What is a quad mesh? What is a base complex?

\section{Quasi-structured quadrilateral meshing} \label{sec:quadqs}
\vspace{.3cm}

% Pas besoin de rentrer dans les détails : juste citer le papier de Maxence
The first step of our pipeline is to obtain
a high-quality unstructured quadrilateral mesh
from the given CAD model.
A mesh vertex is \emph{regular} if its valence (number of incident faces)
is equal to 4; otherwise it is \emph{irregular}.
A quad mesh is said to be \emph{quasi-structured}
if it has few irregular vertices
compared to meshes produced with usual unstructured techniques.
Our group has recently proposed a robust end-to-end pipeline
to achieve this task~\cite{reberol2021}, which is freely available
in the open-source software Gmsh \cite{gmsh}.
The idea is to use a boundary-aligned smooth cross field
to guide the point insertion in a frontal mesher,
then to improve the quad mesh topology by remeshing large convex cavities
while respecting the cross field singularities.
This method forms the first step of our pipeline and is used to generate
the initial unstructured quad mesh.

\begin{figure}[htbp]
    \centering
    \begin{subfigure}{0.33\linewidth}
        \centering\includegraphics[width=\linewidth]{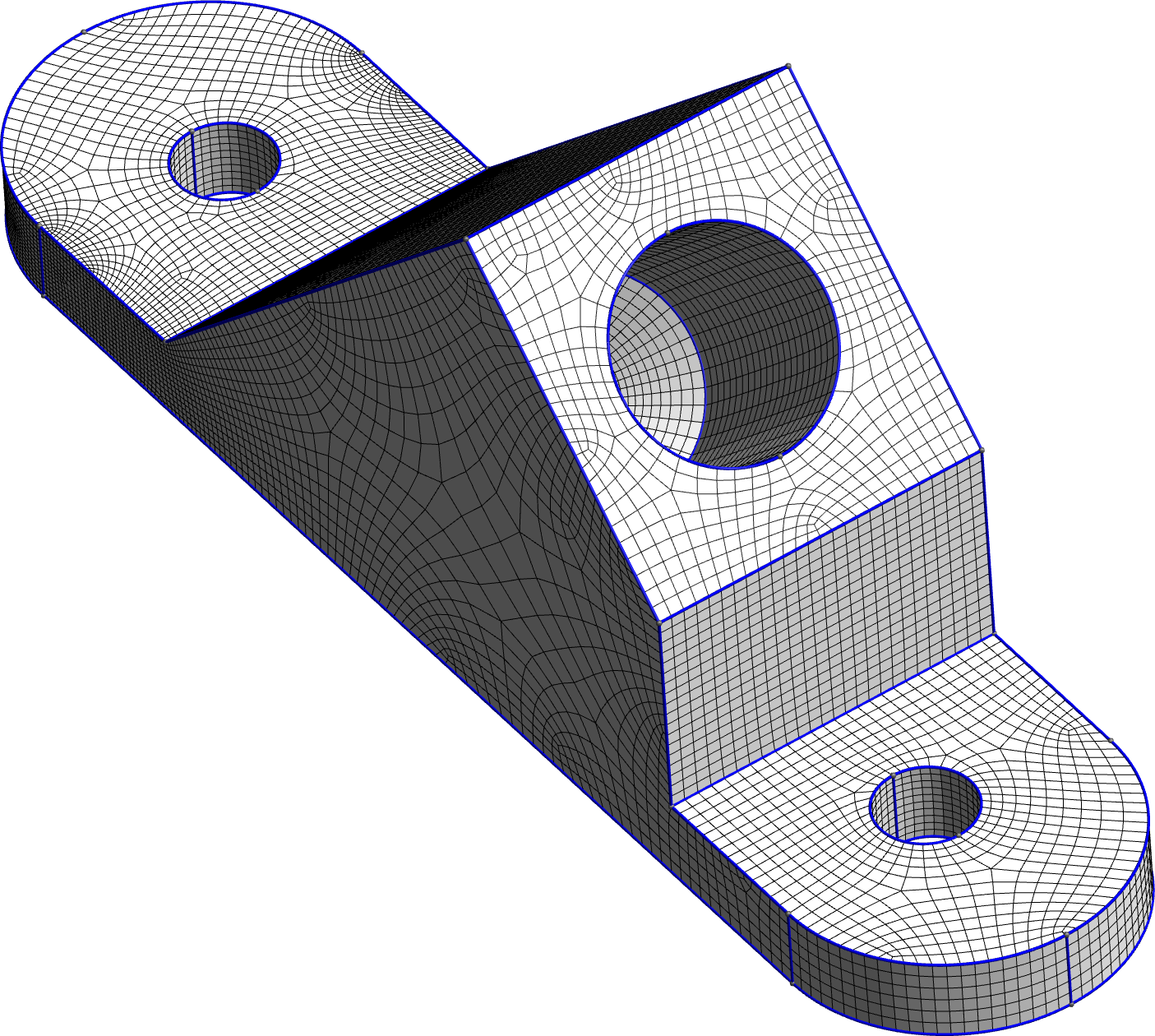}
        \caption{Unstructured mesh}
    \end{subfigure}
    \hspace{0.1\linewidth}
    \begin{subfigure}{0.33\linewidth}
        \centering\includegraphics[width=\linewidth]{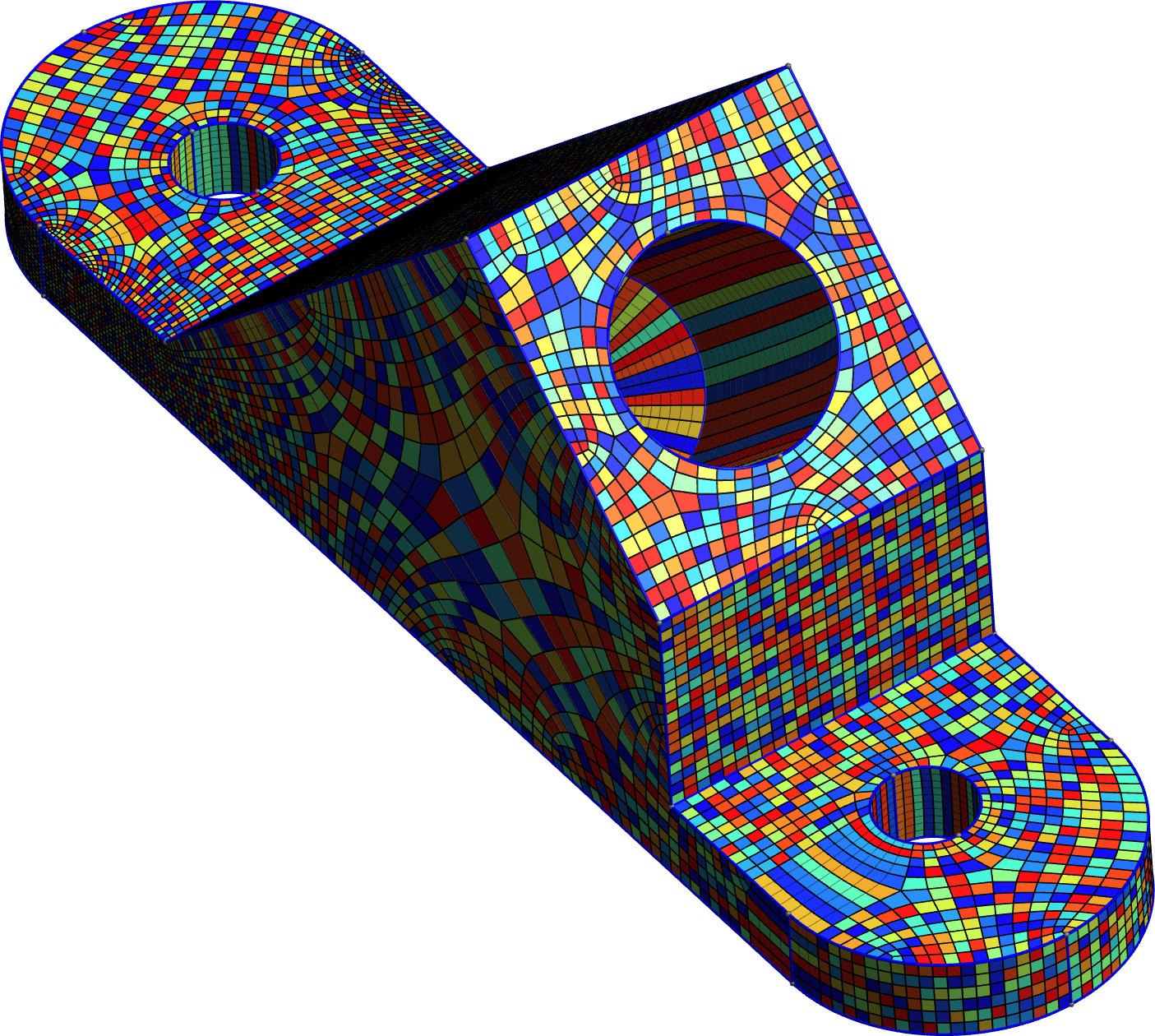}
        \caption{Corresponding quad layout}
    \end{subfigure}
    \caption{Quadrilateral layout of an unstructured mesh
    generated by our quasi-structured mesher.
    The quad layout is very dense. CAD model is M2 from MAMBO dataset~\cite{mambo}.}
    \label{fig:input_layout}
\end{figure}

Even though we are able to produce good linear quadrilateral meshes
(in terms of number of irregular vertices and quad shapes),
they cannot be coarsened directly due to the density of their \emph{quad layout}.
The quad layout is a minimal partition of the mesh
into structured quadrilateral \emph{patches},
obtained by propagating separatrices from the irregular vertices
until they reach other irregular vertices.
A dense layout is illustrated in~\autoref{fig:input_layout}: very few
patches of the layout have more than one single quad of the initial
mesh so the number of patches in the layout is about the same as the number of
quads in the initial mesh.
There are two main reasons for the quad layout to be very dense
for our quasi-structured meshes:
\begin{itemize}
    \item Irregular vertices are not correctly \emph{aligned} on the mesh,
    causing most separatrices to wind around the model
    until meeting an irregular vertex potentially far away.
    \item Many irregular vertices do not match singularities from the cross
    field, but are present to accomodate size transitions and
    small CAD features. They are typically grouped in valence 3-5 pairs.
    These irregular vertices are not all necessary to produce
    good high-order quadrilateral patches.
\end{itemize}
In the next section, we will detail a methodology that allows to
correct these defects and produce a mesh with a coarse
quadrilateral layout.

\section{Quadrilateral layout simplification} \label{sec:layout}
\vspace{.3cm}

% Goal and approach (cite paper)
Given a \emph{quasi-structured} quadrilateral mesh (the \emph{input mesh}), 
we wish to obtain a \emph{block-structured} mesh,
having few irregular vertices but these vertices
should be \emph{aligned} to be able to extract coarse quadrilateral patches.
To this end, we propose a slightly modified version of a method
developed by Lyon~et~al.~\cite{lyon2021},
adapted to incorporate strict compliance to the CAD geometry.
In this section, we present our adaptation of the method.

% % Types of irregular vertices, and explain which we want to keep
% Every irregular vertex in the input mesh can be classified
% in one of the following categories:
% \begin{enumerate}
%     \item Irregular vertices matching CAD corners.
%     \item Irregular vertices corresponding to cross field singularities.
%     These vertices are given by the quasi-structured mesher.
%     \item Remaining irregular vertices serve as size transitions
%     and come in valence 3-5 pairs. These vertices
%     are non-essential and will be removed in the resulting layout.
% \end{enumerate}
% Irregular CAD vertices (1) must remain present in the final mesh
% to have CAD compliance.
% Similarly, cross field singularities (2) are necessary to guarantee
% the topological validity.
% However, remaining irregular vertices (3) are not strictly necessary
% to produce a valid quad mesh, and to achieve our coarseness goal
% we will allow every 3-5 pair to collapse,
% which will remove both irregular vertices.

% How we will enforce geometrical consistency

% Objective

% Not all CAD curves are part of the quadrilateral layout
% such as previously defined:
% some of them only serve to separate neighboring CAD surfaces.
% Still, CAD compliance requires that the resulting quadrilateral
% patches are aligned with all CAD curves.
% Hence, the quadrilateral layout that is considered
% is defined as the \emph{union} of
% the separatrices emanating from irregular vertices,
% and the set of CAD curves.

The objective of the optimization is to produce a layout
that is both coarse and consistent with the geometry, i.e.,
with patches that are not too distorted.
There is a trade-off between these two criteria:
the coarsest layout will have too much distorsion,
and the layout with the best geometry
is the layout of the input mesh, which, as previously shown, is very dense.
To constrain the distorsion of the patches,
we prevent two irregular vertices from aligning
if the correponding deviation on the input mesh
is larger than some prescribed angle $\alpha$.

% The objective of the optimization is to produce a mesh
% with the coarsest possible quad layout,
% i.e., a minimal number of patches.
% In order to keep geometrical consistency,
% we prevent pairs of irregular vertices from aligning
% if their misalignment is too large.
% \note{coarsest may be too distorted. There is a trade-off. We want coarsest with good patches. Thus
% not align everything}

% Overview of the method
The approach works as follows.
First, a \emph{T-mesh}, i.e., a quadrilateral mesh with T-junctions,
is constructed on the given quasi-structured mesh.
Every edge of this T-mesh, or \emph{arc}, has an integer length,
corresponding to the number of mesh edges lying on this arc.
This integer length assignment is known as a \emph{quantization}
of the T-mesh.
The idea of the method is to change this quantization
in order to optimize the quadrilateral layout.
In practice, this will involve assigning many arcs a length of zero.
We formulate this task as an \emph{integer linear program} (ILP),
i.e., an optimization problem with integer variables
and linear objective function and constraints.
On typical models such as the ones treated in this work, the ILP is large, with
thousands of variables and constraints. But such problems can be solved
efficiently thanks to their linearity.

\subsection{T-mesh construction}
From every irregular vertex we propagate \emph{traces},
as if we were drawing the mesh's quad layout.
However, the trace will not be extended until meeting another irregular vertex,
but instead stop at some point when reaching another trace, creating a T-junction.
Such a construction is sometimes called a \emph{motorcycle graph} in reference to the movie Tron.
Contrary to a classical motorcycle graph, traces do not stop immediately
when reaching another trace.
Instead, a trace should stop when it has encountered enough candidates for alignment;
specifically, when it has met one candidate on each side of the trace
that does not deviate by an angle larger than $\alpha$.
Given a path between two vertices $i$ and $j$, $\tan(\alpha_{ij})$ is the ratio
$\frac{l_{ji}}{l_{ij}}$ between the integer length in the orthogonal direction
($l_{ji}$) and in the tangential direction ($l_{ij}$) \cite{lyon2021}.
Formally, we draw a fictitious cone around the trace of half-angle $\alpha$.
The trace stops when it has intersected with two orthogonal traces
on each side
whose origin lies inside the cone.
The rationale is the following:
we want a trace to go far enough to have a couple of potential
irregular vertices to align with,
but we do not want it too long to prevent aligning irregular vertices
far from each other, complexifying the quad layout in vain.

% To determine where the trace stops, we draw a fictitious cone around it with half-angle $\alpha$. 
% Our problem formulation will prevent two irregular vertices
% from aligning if they do not lie in each other's cone.
% The angle $\alpha$ is prescribed by the user,
% allowing him to choose the geometrical consistency
% he wishes to enforce: a small angle will lead to a high number
% of patches, and vice-versa.
% A trace stops when it has met, on both sides,
% another irregular vertex lying inside the cone.
% The rationale is the following:
% we want a trace to go far enough to have a couple of potential
% irregular vertices to align with,
% but we do not want it too long to prevent aligning irregular vertices
% far from each other, complexifying the quad layout in vain.

In order to comply with the CAD model,
it is forbidden for non-CAD irregular vertices
to collapse on a CAD curve.
A simple way to address this is to set $\alpha = 0$
for traces lying on CAD curves.
This will prevent the trace from diverting towards
an irregular vertex away from the CAD curve.
Note that, according to the stopping rule, a trace lying on a CAD curve will
not stop until meeting another irregular vertex.

The constructed T-mesh $ \mathcal{T} = (\mathcal{N}, \mathcal{A}, \mathcal{P}) $
consists of nodes $\mathcal{N}$ for every irregular vertex and intersection of traces,
arcs $\mathcal{A}$ corresponding to segments of a trace between two nodes,
and patches $\mathcal{P}$ which are the quadrilateral regions bounded by the arcs.
Every arc $ a \in \mathcal{A} $ is assigned an integer length $ q_a $;
the set of integer lenghts are the variables of our optimization problem.

\begin{figure}
    \begin{center}
        \includegraphics[width=\textwidth]{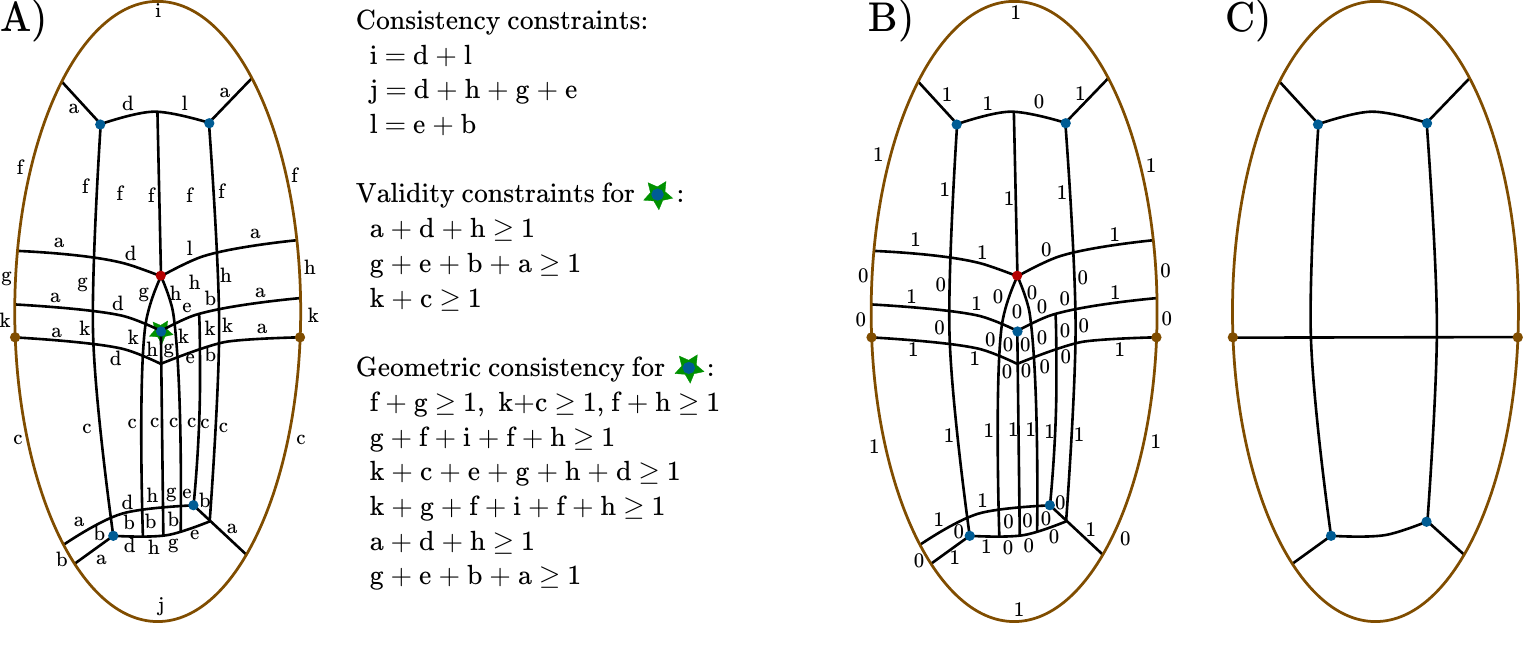}
        \caption{Illustration of the simplification of a T-mesh quantization.
            From the quad mesh of an ellipse, the T-mesh (A) is built, with
            irregular vertices of valence three in blue, valence five in red
            and CAD corners in brown. Alphabet letters are the integer lengths
            in the T-mesh quantization. After minimizing the objective function
        (weighted sum of lengths) subject to the linear constraints, we
    obtain the quantization in (B), where many arcs are collapsed (length
equals to zero). The equivalent quad layout is coarse (C).}
        \label{fig:ellipse}
\end{center}
\end{figure}

\subsection{Quantization constraints}
To achieve our goal of obtaining a coarse quantization
that is valid and geometrically consistent,
several constraints are enforced.

\paragraph{Non-negativity}
It is clear that every arc should have a non-negative length:
for every arc $ a \in \mathcal{A} $,
\begin{equation}
    q_a \geq 0.
\end{equation}

\paragraph{Consistency}
The quantization is said to be \emph{consistent}
if for every patch, pairs of opposite sides
have the same total length.
This ensures that the quantization implicitly defines a conforming quad mesh.
For every patch with (ordered) sides $S_1, S_2, S_3, S_4$,
each side being a set of arcs, we write
\begin{equation}
    \sum_{a \in S_1} q_a = \sum_{b \in S_3} q_b, \quad \text{and} \quad
    \sum_{a \in S_2} q_a = \sum_{b \in S_4} q_b.
\end{equation}

\paragraph{Validity}
For the quantization to be \emph{valid}, irregular vertices
must be separated by a strictly positive distance.
This may seem complicated to formulate,
but luckily the authors of~\cite{lyon2021}
have proved a simple sufficient condition.
Let $i$ be an irregular vertex and $ t_i $ one of its traces
and let $t_*$ denote the first trace that $ t_i $ intersects
whose origin $*$ lies in the $\pi/4$ cone around $t_i$.
Let $ n_{i*} $ denote the intersection,
and $ S_{i*} $ the set of arcs from $i$ to $n_{i*}$.
Then, for an irregular vertex $i$ to be separated from any other irregular vertex,
it is sufficient that $i$ and $ n_{i*} $ are separated.
Therefore, for every trace $t$ emanating from an irregular vertex,
we set the constraint
\begin{equation}
    \sum_{a \in S_{i*}} q_a \geq 1.
\end{equation}
As explained previously, the input mesh usually possesses
many pairs of 3-5 irregular vertices that serve to satisfy size transitions.
To further simplify the layout, we make an exception
to the validity constraint to allow 3-5 pairs to collapse on each other.
Such a collapse is valid as it results in a regular vertex.
In practice, the constraint is not imposed if $i$ and $n_{i*}$
form a 3-5 pair.

\paragraph{Geometric consistency}
In order to ensure consistency with the geometry
defined by the input mesh,
we must explicitly separate irregular vertices
whose traces intersect, but that do not lie in each other's cone.
To do so, for every irregular vertex $i$, every irregular vertex $j$
that is outside the cone of $i$ is separated:
\begin{equation}
    \text{for every $ n_{ij} $ with $ \frac{l_{ji}}{l_{ij}} > \tan \alpha_i $,} \qquad
    \sum_{a \in S_{ji}} q_a \geq 1,
\end{equation}
where $n_{ij}$ is the intersection of $t_i$ and $t_j$,
$ l_{ji} $ is the total length from $j$ to $n_{ij}$,
$ S_{ji} $ is the set of arcs from $j$ to $n_{ij}$,
and $\alpha_i$ is the cone half-angle for trace $t_i$
(which is either the user-prescribed angle $\alpha$,
or 0 is the trace lies on a CAD curve).

\subsection{Objective function}
To obtain the coarsest possible quadrilateral layout,
we need to minimize the number of patches.
Let $a,b$ denote the integer dimensions of some patch.
If $ a = 0 $ or $ b = 0 $, the patch vanishes
and does not contribute to the total number of patches.
If $ a = b = 1 $, we add a single patch to our objective function.
If $ a > 1 $ or $ b > 1 $, it means that there remains T-junctions.
To obtain a conforming quad layout, we split the patch
into a grid of $ a \times b $ patches, each having dimensions $ 1 \times 1 $
and contributing as a single patch to the objective function.
Hence, the objective that exactly promotes coarseness
is the total integer area of the mesh:
\begin{equation}
    \min_q \sum_{p \in \mathcal{P}} \mathrm{area}(p)
    = \sum_{p \in \mathcal{P}} \qty( \sum_{a \in S_1} q_a \times \sum_{b \in S_2} q_b ).
    \label{eq:exact_obj}
\end{equation}
However, this objective is nonlinear,
and using it would not provide us an ILP.
Instead, we minimize a \emph{weighted sum of arc lengths}.
The weight serves to help the solver prioritize arcs to collapse.
To this end, we set the weight of an arc $a$
to be the inverse of its initial integer length $l_a$.
This gives the following objective function:
\begin{equation}
    \min_q \sum_{a \in \mathcal{A}} \frac{1}{l_a} q_a.
\end{equation}

Note that this optimization problem is guaranteed to have a solution since the
input mesh verifies every constraint.

The previous T-mesh quantization simplification problem is illustrated on
Figure \ref{fig:ellipse}, where we build the T-mesh of the quad mesh of an
ellipse, which is not optimal because the irregular vertices are not aligned
and there is an interior 3-5 pair. We list all the consistency constraints, and
the validity and geometric consistency constraints associated to the central
irregular vertex of valence three. Note that by using a single variable (e.g.
length \emph{a} along the ellipse) for multiple arcs, we already enforce
the simple equality consistency constraints. In this particular example, the
validity constraints are redundant with some of the geometric consistency
constraints, but this is not always the case. The solution of the integer
problem (values in B) contains lengths equal to zero and one, where
zero corresponds to an arc collapse. By merging the appropriate vertices
and smoothing the resulting quad mesh, we obtain the coarse layout in C.
The horizontal line is kept because it connects the CAD corners (dots in brown),
which are preserved in our formulation.

\section{Block-structured quad mesh and high-order quadrilaterals}
\vspace{.3cm}
\label{sec:bsquad}

The optimized quantization implicitly defines a topological quad layout.
In this section, we explain how we obtain a smooth and fine block-structured mesh.
First, a conforming coarse quad mesh is extracted from the quantization;
see \ref{ssec:splitting}, \ref{ssec:merging}, \ref{ssec:placement}.
However this mesh has badly placed vertices and non-straight edges.
Our approach to correct this is to subdivide the coarse patches
to obtain a fine mesh; see~\ref{ssec:subdivision},
then smooth the fine mesh to optimize the placement of vertices
and straighten the edges; see~\ref{ssec:smoothing}.

\subsection{Edge and patch splitting} \label{ssec:splitting}
In the optimized quantization,
some arcs will be assigned an integer length $ q \geq 2 $.
These edges are split into $ q $ sub-edges.
To obtain a conforming quad layout,
this split must be "propagated" by splitting the incident patches:
if an incident patch has integer dimensions $ q \times q' $,
and $ q' > 0 $, it is split into $ q \cdot q' $ sub-patches
of dimensions $ 1 \times 1 $;
if $ q' = 0 $ it is split into $ q $ sub-patches of dimensions $ 1 \times 0 $.
Splitting patches of zero area is necessary to
correctly merge vertices in the next step.

\subsection{Vertex and edge merging} \label{ssec:merging}
At this point we have a conforming layout
but having many zero-length edges and zero-area patches.
To get rid of these,
we first identify groups of vertices that have zero distance between each other
(formally, two vertices are in the same group if they are connected
on the T-mesh by a path of zero total length).
Every group of vertices is virtually merged into a single "center vertex".
This effectively removes all zero-length T-edges.
Then, edges bounded by the same pair of center vertices
are virtually merged into one edge.
This removes all zero-area patches,
and leaves us with a clean topological mesh.

\subsection{Vertex and edge placement} \label{ssec:placement}
Vertices and edges need to be assigned a geometrical location on the model.
This operation requires to pay attention to CAD consistency.
Center vertices are chosen in the following way:
\begin{itemize}
    \item If some vertex in the group corresponds to a CAD vertex,
    the latter is taken as the center vertex
    (there is at most one CAD vertex in the group);
    \item Otherwise, if some vertex in the group lies on a CAD curve,
    any vertex on the curve is taken as the center vertex
    (there is at most one such curve in the group);
    \item Otherwise, any vertex of the group is taken as center.
\end{itemize}
Afterwards, edges are formed by taking the shortest path on the input mesh
between the two centers they connect.
If an edge lies on a CAD curve
(i.e., some of the T-edges before the merge lied on a CAD curve),
its corresponding path is the subset of the curve that connects the two centers.

\subsection{Patch subdivision} \label{ssec:subdivision}
Once we have a conforming quad mesh,
we subdivide the coarse patches to refine the mesh.
Note that opposite sides of a patch must have the same subdivision.
Therefore, an integer width has to be chosen for every so-called "quad loop".
A quad loop is a closed sequence of patches obtained by starting from
a patch and iteratively go to its neighboring patch, following the same direction.

A target integer length is assigned to every edge,
defined as the length of its corresponding path on the input mesh.
Since this target length is not constant along the quad loop,
the (rounded) mean target length is taken as the quad loop's integer width.
Following this process for every quad loop will assign to every edge an integer length.
Finally, edges are subdivided into equal parts,
and patches are subdivided by transfinite interpolation.

\subsection{Smoothing} \label{ssec:smoothing}
Once we have a fine mesh, we can proceed to smooth the mesh.  We perform
Winslow smoothing~\cite{winslow1966} which is the industry-standard technique
for structured grid mesh generation.  In an explicit optimization loop, each
vertex is moved according to its neighbors.  The Winslow coefficients are
computed with a finite difference discretization of the Winslow equation
\cite{knupp1999}. Since this operation moves vertices away from the CAD
surface, every vertex is projected back on it by finding the closest point on
the surface. 

This smoothing tends to create squared quadrilaterals, so it is important that
in the previous patch subdivision (\ref{ssec:subdivision}), the integer
lengths assigned to the quad loops are chosen according the geometry.
After smoothing, the block-structured quad mesh usually contains patches with
rectangular shapes (e.g. Fig. \ref{fig:overview}.d.).

The explicit loop, with the FDM discretization of Winslow equations and
CAD projections, is not guaranteed to produce valid quads (strictly positive
Jacobian). In practice, the behaviour is generally satisfactory, but
negative quads do appear at some concave corners. In the future, we will
explore additional untangling techniques for surfaces, or directly
optimize the high-order quad mesh produced later in the pipeline.

\subsection{Extraction of high-order quadrilaterals} \label{ssec:extraction}

From the smooth block-structured quad mesh, we can extract high-order quadrilaterals.
Assuming the elements are parametrized with Lagrange finite element functions of 
order $N$, the process is straightforward:
\begin{itemize}
    \item For each vertex $v_i$ of the block structure, we create a high-order vertex.
    \item For each edge $(v_i,v_j)$ of the block structure, we retrieve the two
        high-order extremities and we create $N-1$ interior high-order vertices
        along the curved edge via polyline equidistant re-sampling. 
        Each new high-order vertex is projected on the CAD.
    \item For each patch $(v_i,v_j,v_k,v_l)$, we retrieve the four high-order corners and the four 
        high-order sides, and we create $(N-1)^2$ interior high-order vertices with
        transfinite interpolation.  Each new high-order vertex is projected on the CAD.
\end{itemize}

The above process generates high-order meshes with regularly spaced nodes. All the
figures shown in this paper were produced with order five polynomials.

For better numerical properties, it is possible to produce Gauss-Lobatto
spacing by slightly changing the curve re-sampling step. One could also build
B-spline surfaces to better match the initial CAD geometry. In the future, we
plan to implement multiple high-order representations and let the user choose
the one best suited for his use cases.

Note that the position of the high-order nodes is entirely dependant on the result
of the block-structured linear quad mesh smoothing. It may be interesting to employ
high-order surface smoothing to further improve the mesh geometry, but such
endeavour is challenging and we keep it for further work.

\section{Results and discussion}
\begin{figure}[htbp]
    \centering
    \includegraphics[width=0.3\textwidth]{figures/align/M1.png}
    \includegraphics[width=0.3\textwidth]{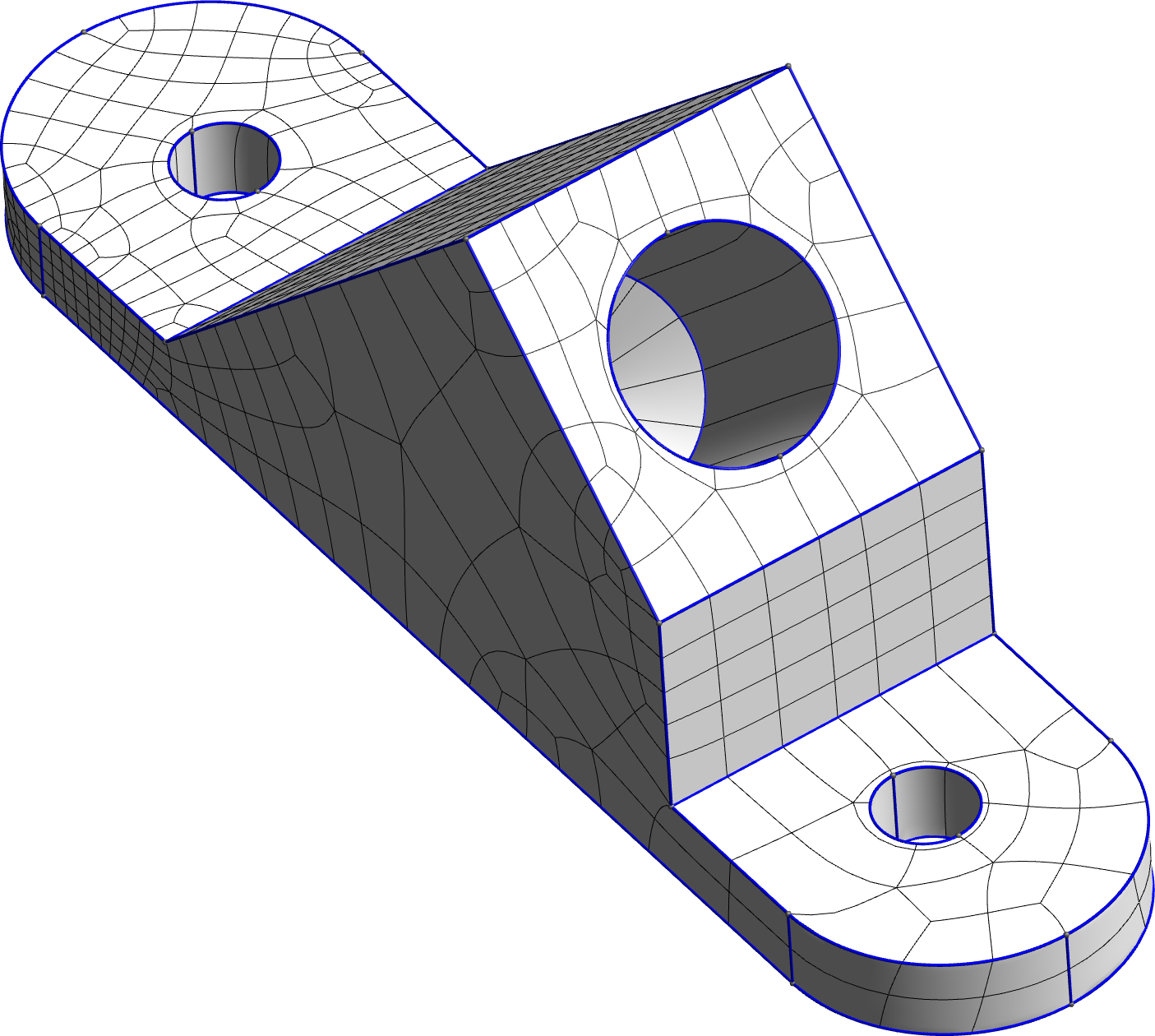}
    \includegraphics[width=0.3\textwidth]{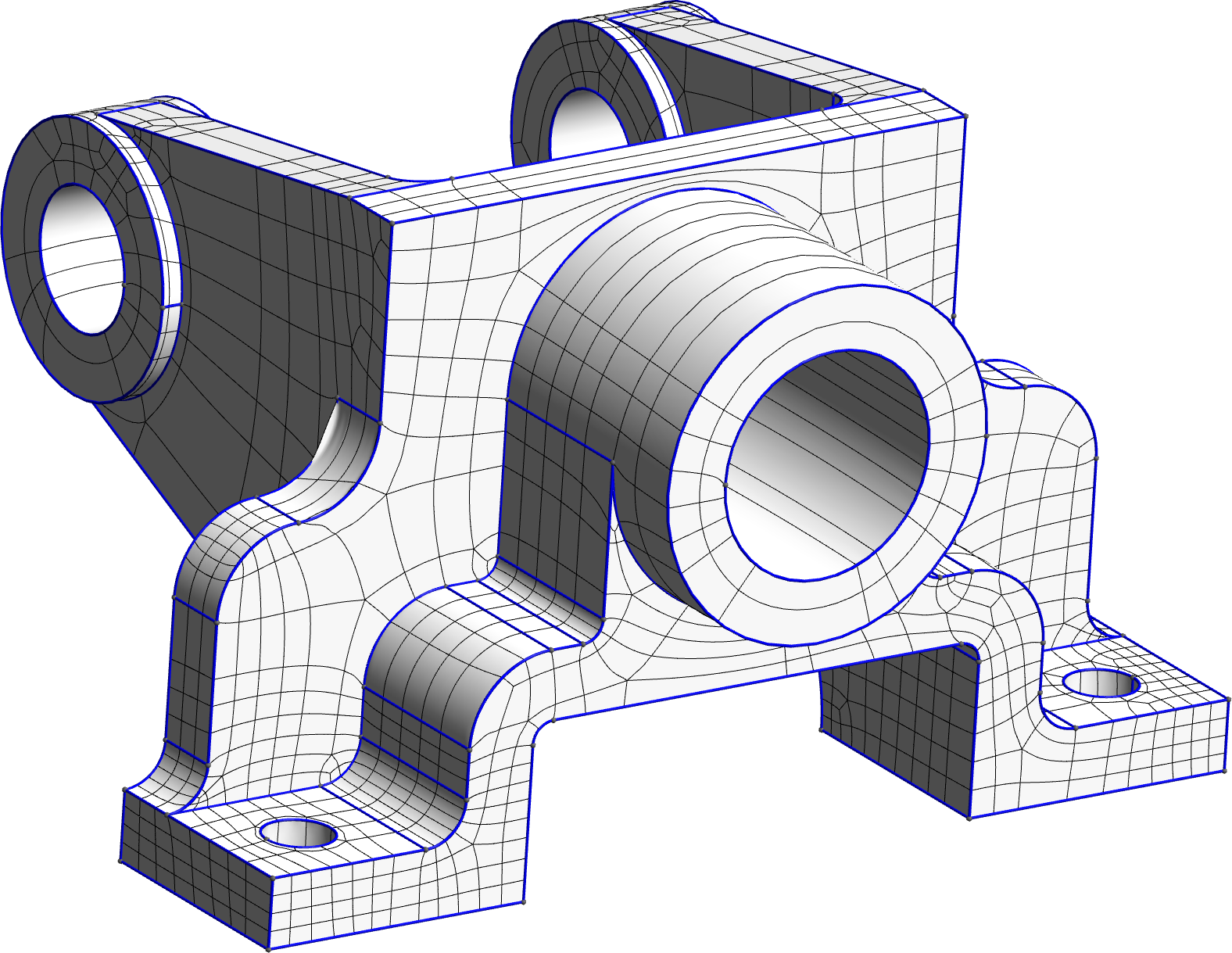} \\
    \includegraphics[width=0.3\textwidth]{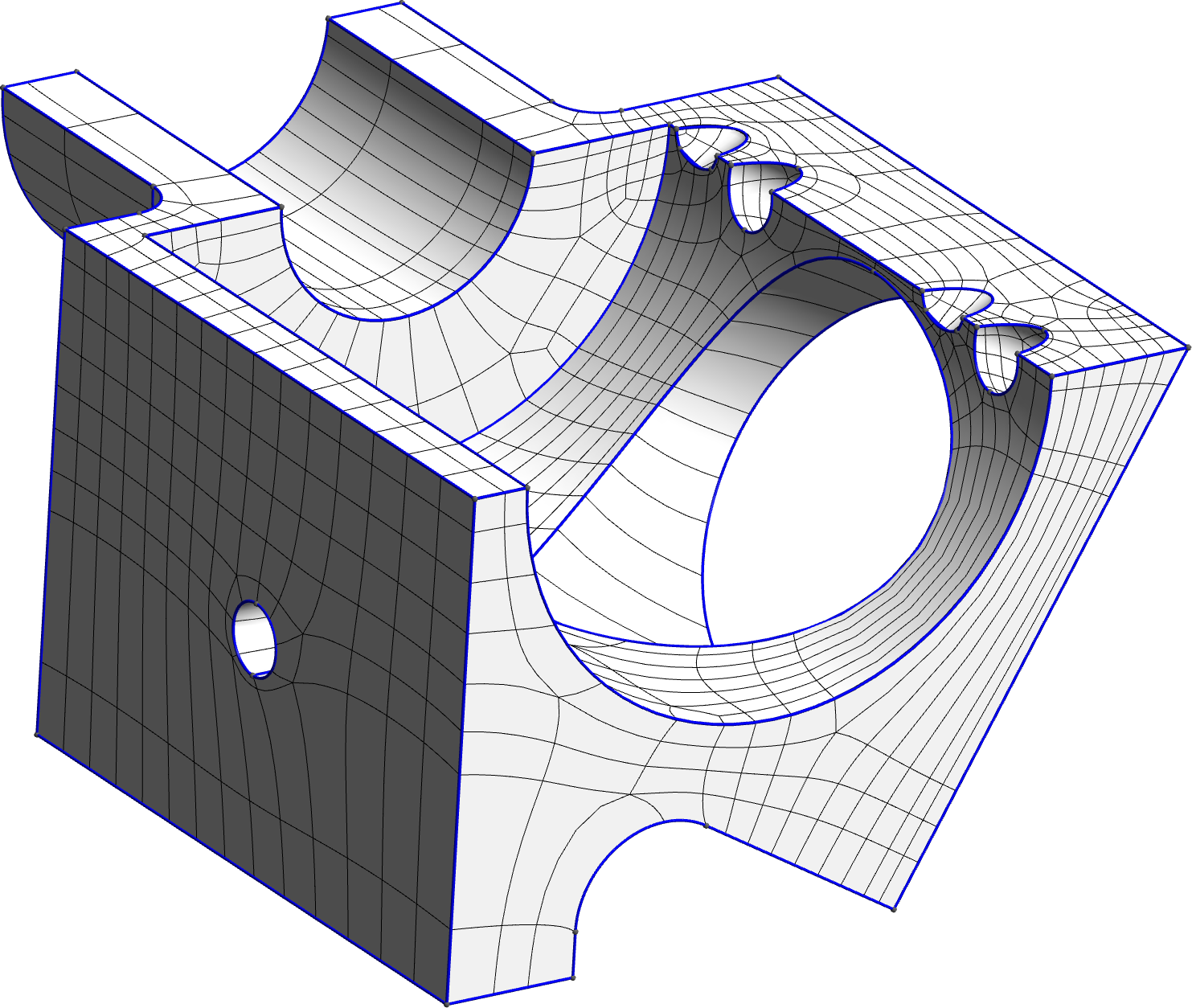}
    \includegraphics[width=0.3\textwidth]{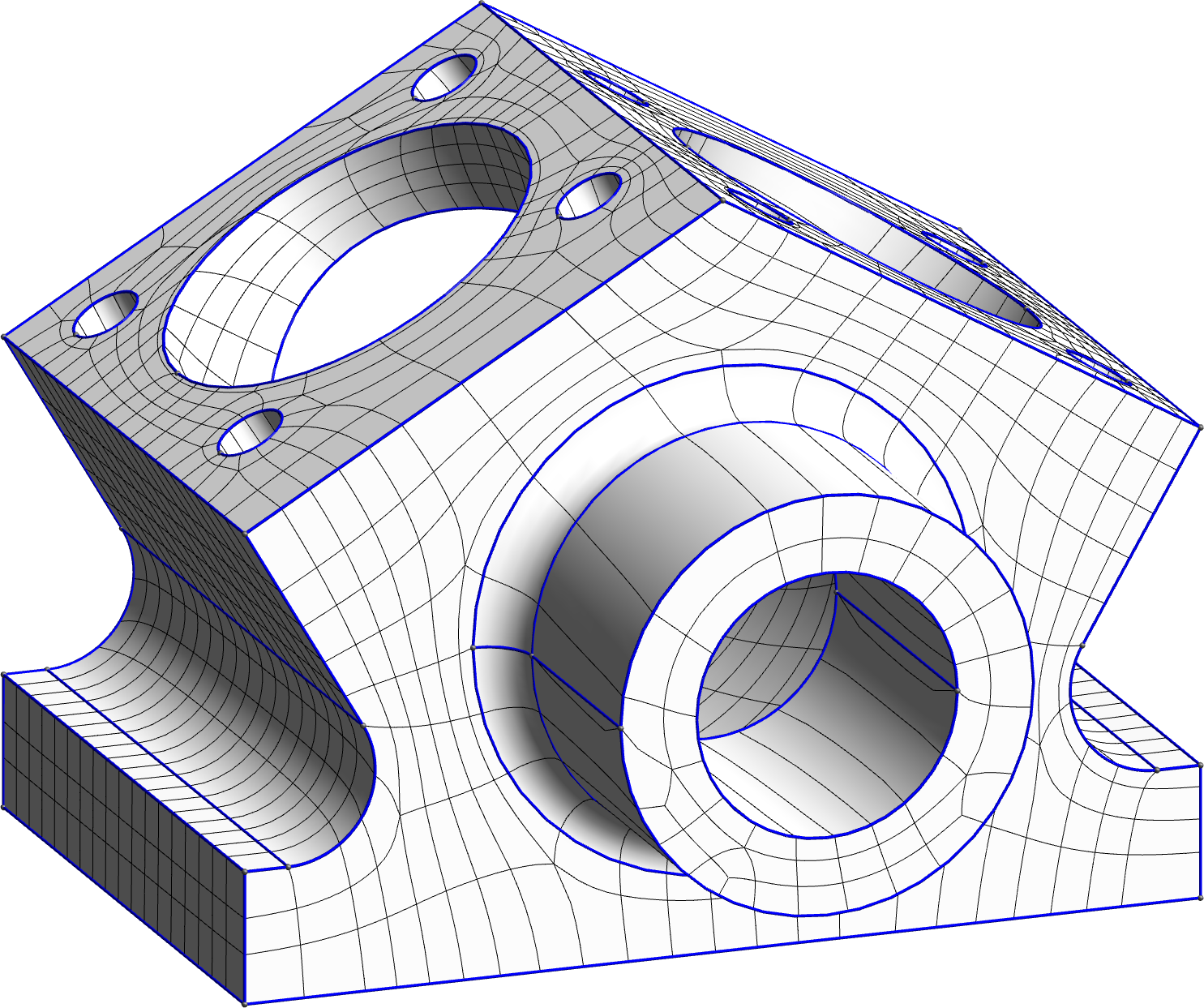}
    \includegraphics[width=0.3\textwidth]{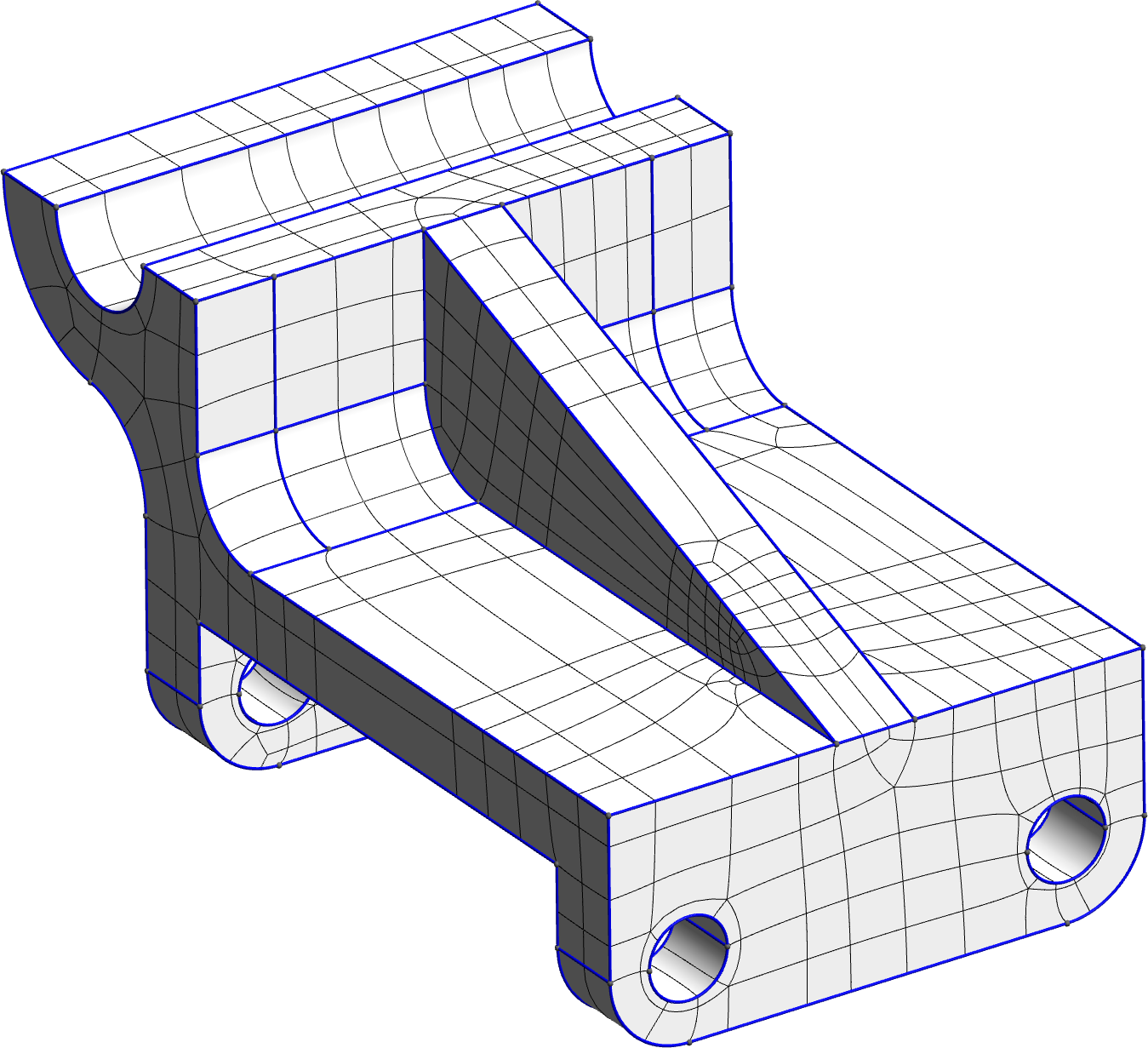} \\
    \includegraphics[width=0.5\textwidth]{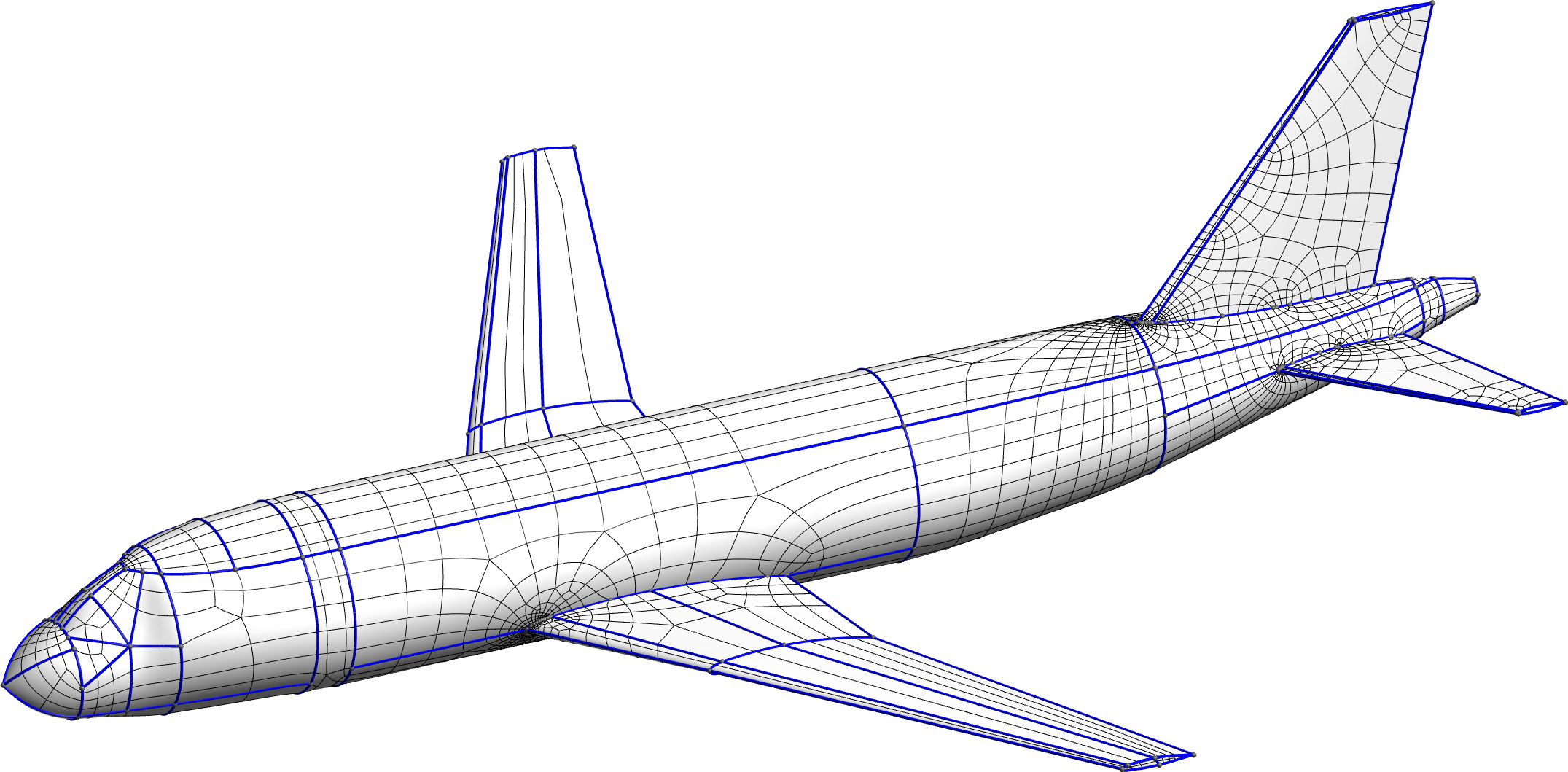}%
    \includegraphics[width=0.5\textwidth]{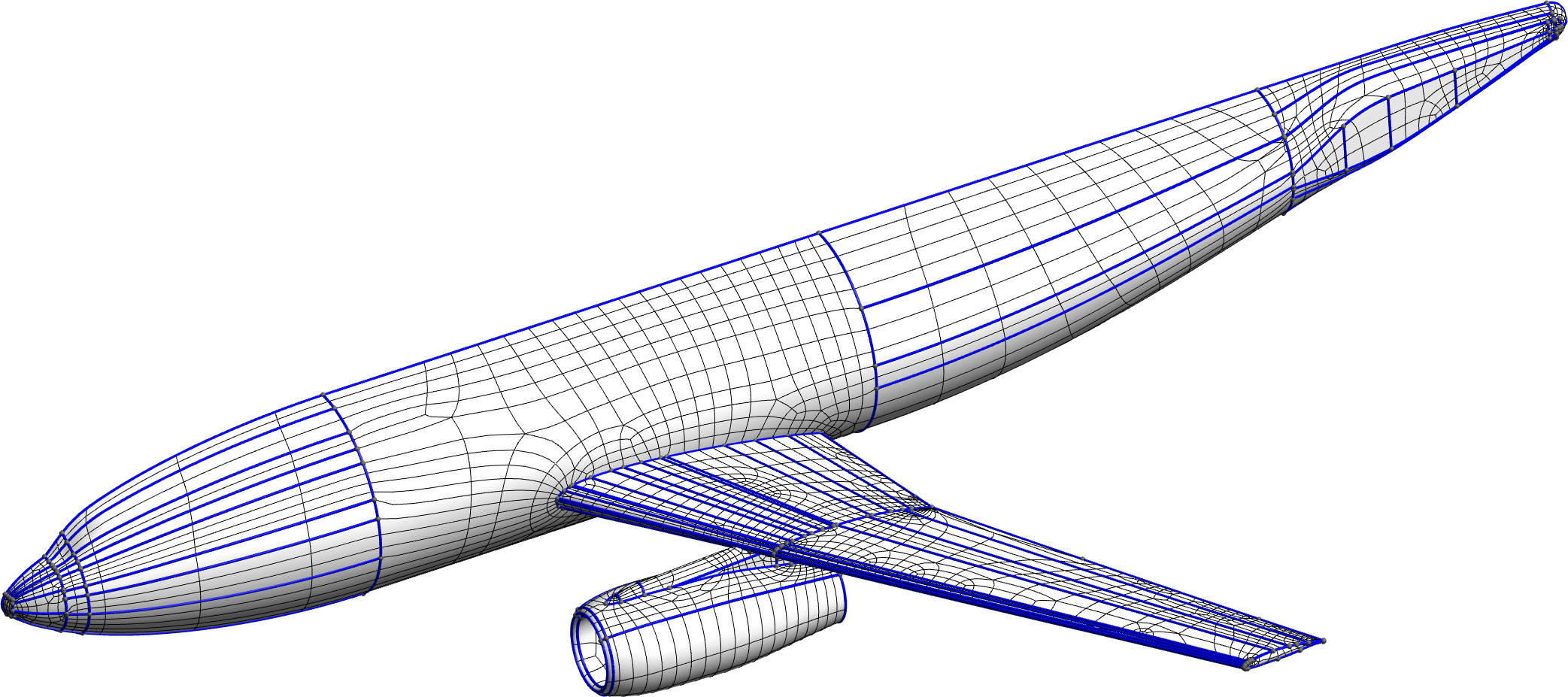}
    \caption{Results of our method on a variety of CAD models
    (5th-order quadrilaterals, $ \alpha = \pi/4 $).
    Top six: various models from the MAMBO dataset~\cite{mambo} (M1, M2, M3, M4, M5, M6).
Bottom two: Airbus A319 and DLRF6~\cite{brodersen2002}.}
    %\note{A319 source?}
    \label{fig:results}
\end{figure}

\begin{table}[htbp]
    \centering
    \begin{threeparttable}[b]
        \caption{Statistics of results in \autoref{fig:results}.}
        \label{tab:stats}
        \begin{tabular}{|l|rrr|rr|rrrr|}\hline
            \textbf{Model} & \textbf{\#Faces}\tnote{1} & \textbf{\#P$_\mathrm{init}$}\tnote{2} & \textbf{\#Irr}\tnote{3} & \textbf{\#Vars}\tnote{4} & \textbf{\#P$_\mathrm{final}$}\tnote{5} & \textbf{Time}\tnote{6} & \textbf{QuadQS}\tnote{7} & \textbf{ILP}\tnote{8} & \textbf{Smooth}\tnote{9} \\\hline
            \begin{filecontents*}{data/stats.txt}
M1, 23707, 8625, 180, 3340, 1168, 26.89978, 71.71062365565815, 3.557579281317542, 12.929139197420945
M2, 16059, 12253, 156, 3541, 983, 17.290300000000002, 66.56911678802565, 7.477718720901314, 12.614703041589792
M3, 22242, 15864, 336, 5968, 2505, 28.32101, 68.64161977274117, 11.177002515093918, 9.249458264376871
M4, 25996, 15394, 194, 3702, 1596, 29.5512, 61.58802349819973, 4.2306911394461135, 22.923637618776905
M5, 32002, 20487, 287, 5265, 2681, 45.2299, 66.34991454767753, 5.353538256772622, 18.25944784313032
M6, 26322, 19975, 118, 2522, 1007, 31.2675, 44.742943951387225, 1.788192212361078, 42.90653234188855
A319, 18193, 13616, 410, 7565, 4278, 173.4093, 35.1280467656579, 3.5430971695289695, 53.201875562614006
DLRF6, 99978, 42354, 620, 11830, 6553, 282.36990000000003, 29.587749969100805, 5.630593062504183, 57.38394920988391
\end{filecontents*}
\csvreader[no head, late after line=\\\hline]{data/stats.txt}%
            {1=\model,2=\faces,3=\pinit,4=\irr,5=\vars,6=\pfinal,7=\time, 8=\quadqs, 9=\ilp, 10=\smooth}%
            {\model & \faces & \pinit & \irr & \vars & \pfinal & \SI[round-mode=places, round-precision=0]{\time}{\second} & \SI[round-mode=places, round-precision=0]{\quadqs}{\percent} & \SI[round-mode=places, round-precision=0]{\ilp}{\percent} & \SI[round-mode=places, round-precision=0]{\smooth}{\percent}}
        \end{tabular}
        \begin{tablenotes}
            \item[1] Number of faces in quasi-structured mesh.
            \item[2] Number of patches in the quasi-structured mesh layout.
            \item[3] Number of irregular vertices in quasi-structured mesh.
            \item[4] Number of variables in the ILP.
            \item[5] Number of patches in high-order mesh.
            \item[6] Total run time from CAD to high-order mesh.
            \item[7] Proportion of time used for quasi-structured meshing.
            \item[8] Proportion of time used for solving the ILP.
            \item[9] Proportion of time used for smoothing and building the high-order mesh.       
        \end{tablenotes}
    \end{threeparttable}
\end{table}

\vspace{1em}
To validate our approach and its robustness,
we apply it on several existing CAD models,
and discuss its strengths and weaknesses.

\paragraph{Implementation}
The complete pipeline is implemented in our meshing software Gmsh.
The quasi-structured quadrilateral mesher and Winslow smoother
are readily available in Gmsh.
For solving the ILP we use \texttt{lp\_solve}
which is a free and open-source Mixed-Integer Linear Programming solver~\cite{lp_solve}.
In our experiments, the complete pipeline is run on a single core
on an average laptop computer.

\paragraph{CAD models}
For illustration, we have chosen six CAD models from the MAMBO dataset~\cite{mambo} (M1 to M6) as
well as two aircraft models, an Airbus A319 and the
DLR-F6~\cite{brodersen2002}.  As a maximal deviation we chose $ \alpha = \pi/4
$ as we found that it achieves a good balance between coarseness and mesh
quality.  The order of the quadrilaterals is $ N = 5 $.

On~\autoref{fig:results} we illustrate
the high-order meshes produced by our method
on the given CAD models.
In~\autoref{tab:stats} we provide some statistics 
about the process.

\paragraph{Discussion}
First we can notice that the total run time, from a few seconds to a couple of minutes,
is reasonable in an engineering analysis context,
even on the more complex aircraft models.
In all cases, solving the integer optimization takes a very small portion of 
the total run time.
From this observation we are confident that ILP formulations
are a viable approach to high-quality, high-order mesh generation.
As the model complexity increases,
smoothing takes a significant portion of the time.

Our method is able to significantly reduce the complexity of the quad layout,
as can be deduced from the $ \#P_\mathrm{init} / \#P_\mathrm{final} $ ratio 
which ranges from $3$ to $11$.
While the mesh coarseness is satisfying,
we believe that it can be further simplified without
affecting significantly the geometric consistency.
We see two limitations that prevent a better simplification:
\begin{itemize}
    \item Once a 3-5 pair of irregular vertices has been collapsed
    into a regular vertex, this vertex cannot be collapsed further
    due to the validity constraints enforced on the former irregular vertices.
    \item After quantization,
    many arcs have an integer length greater than 1,
    which significantly increases the number of patches after the splitting phase.
    This is due to our choice of objective function
    which is a linearization of the total mesh area.
\end{itemize}

By visual inspection we see that the quadrilaterals have low distortion,
and while being coarse, they accurately capture the model's geometry.
The meshes we have produced are therefore perfectly suitable for high-order analysis techniques.

\section{Conclusion and Future Work}
\vspace{1em}
In this work, we have presented a high-order meshing pipeline
that is fast, effective and robust with respect to
the CAD model in input.
It combines our robust quasi-structured quad mesher
together with a layout simplification based
on an ILP formulation that allows to flexibly impose CAD compliance.
An additional advantage is that the layout simplification
only requires an unstructured quad mesh as input;
there is no need for a seamless parametrization or a cross field
which are usually hard to build robustly on CAD models.
Although we have not achieved industrial-grade performance
in terms of computational cost and coarseness,
we are confident that further improvements can,
paving the way for high-order analysis
on arbitrarily complex CAD geometries.

In terms of \textbf{computational cost},
we should concentrate our efforts on the smoothing step
which takes a large portion of the total time 
on large models.
Another clue we would like to investigate is parallelism:
a first "pre-simplification" could act on every CAD face
separately while enforcing a quantization on the boundaries,
after which a global layout simplification would finish the work.
% \note{how to improve smoothing?}

% we noticed that the time for solving the ILP
% does not scale well with the size of the problem.
% This could be improved by using a more sophisticated MILP
% solver that is more scalable.

In terms of \textbf{coarseness},
as previously said, we believe further
simplification can be achieved without sacrificing geometric consistency.
A first improvement is to allow collapsed 3-5 pairs to further collapse
onto model curves or other irregular vertices.
This could be achieved, e.g., using conditional constraints.
Furthermore, by trading off computational performance,
we could investigate the use of the exact objective function~\eqref{eq:exact_obj}
with non-linear solvers.
Finally, we note that the quality of the CAD model severely impacts
the coarseness that can be achieved.
This is particularly visible on the aircraft models in~\autoref{fig:results}
which possess many short CAD curves.
Relaxing the CAD constraints by correcting such defects
would allow to achieve a coarser result.

% \section*{Appendix}

% \newpage

\section*{Acknowledgments}

This research is supported by the European Research Council (project HEXTREME, ERC-2015-AdG-694020).
Mattéo Couplet is a Research Fellow of the F.R.S.-FNRS.

\bibliography{sample}

\begin{thebibliography}{13}
\newcommand{\enquote}[1]{``#1''}
\providecommand{\natexlab}[1]{#1}
\providecommand{\url}[1]{\texttt{#1}}
\providecommand{\urlprefix}{URL }
\expandafter\ifx\csname urlstyle\endcsname\relax
  \providecommand{\doi}[1]{\discretionary{}{}{}https://doi.org/#1}\else
  \providecommand{\doi}[1]{\discretionary{}{}{}\urlstyle{rm}\url{https://doi.org/#1}}\fi

\bibitem[{Hughes et~al.(2005)Hughes, Cottrell, and Bazilevs}]{hughes2005}
Hughes, T. J.~R., Cottrell, J.~A., and Bazilevs, Y., \enquote{Isogeometric
  analysis: {CAD}, finite elements, {NURBS}, exact geometry and mesh
  refinement,} \emph{Computer Methods in Applied Mechanics and Engineering},
  Vol. 194, No.~39, 2005, pp. 4135--4195.
\newblock \doi{10.1016/j.cma.2004.10.008}.

\bibitem[{Sevilla et~al.(2008)Sevilla, Fern{\'a}ndez-M{\'e}ndez, and
  Huerta}]{sevilla2008}
Sevilla, R., Fern{\'a}ndez-M{\'e}ndez, S., and Huerta, A.,
  \enquote{NURBS-enhanced finite element method (NEFEM),} \emph{International
  Journal for Numerical Methods in Engineering}, Vol.~76, No.~1, 2008, pp.
  56--83.

\bibitem[{Bommes et~al.(2011)Bommes, Lempfer, and Kobbelt}]{bommes2011}
Bommes, D., Lempfer, T., and Kobbelt, L., \enquote{Global {Structure}
  {Optimization} of {Quadrilateral} {Meshes},} \emph{Computer Graphics Forum},
  Vol.~30, No.~2, 2011, pp. 375--384.
\newblock \doi{10.1111/j.1467-8659.2011.01868.x}.

\bibitem[{Tarini et~al.(2011)Tarini, Puppo, Panozzo, Pietroni, and
  Cignoni}]{tarini2011}
Tarini, M., Puppo, E., Panozzo, D., Pietroni, N., and Cignoni, P.,
  \enquote{Simple quad domains for field aligned mesh parametrization,}
  \emph{ACM Transactions on Graphics}, Vol.~30, No.~6, 2011, pp. 1--12.
\newblock \doi{10.1145/2070781.2024176}.

\bibitem[{Razafindrazaka et~al.(2015)Razafindrazaka, Reitebuch, and
  Polthier}]{rafa2015}
Razafindrazaka, F.~H., Reitebuch, U., and Polthier, K., \enquote{Perfect
  {Matching} {Quad} {Layouts} for {Manifold} {Meshes},} \emph{Computer Graphics
  Forum}, Vol.~34, No.~5, 2015, pp. 219--228.
\newblock \doi{10.1111/cgf.12710}.

\bibitem[{Lyon et~al.(2021)Lyon, Campen, and Kobbelt}]{lyon2021}
Lyon, M., Campen, M., and Kobbelt, L., \enquote{{Quad Layouts via Constrained
  T-Mesh Quantization},} \emph{Computer Graphics Forum}, 2021.
\newblock \doi{10.1111/cgf.142634}.

\bibitem[{Reberol et~al.(2021)Reberol, Georgiadis, and Remacle}]{reberol2021}
Reberol, M., Georgiadis, C., and Remacle, J.-F., \enquote{Quasi-structured
  quadrilateral meshing in {Gmsh} -- a robust pipeline for complex {CAD}
  models,} \emph{arXiv:2103.04652 [cs]}, 2021.

\bibitem[{Ledoux()}]{mambo}
Ledoux, F., \enquote{MAMBO dataset,} , ????
\newblock \urlprefix\url{https://gitlab.com/franck.ledoux/mambo}.

\bibitem[{Geuzaine and Remacle(2009)}]{gmsh}
Geuzaine, C., and Remacle, J.-F., \enquote{Gmsh: {A} 3-{D} finite element mesh
  generator with built-in pre- and post-processing facilities,}
  \emph{International Journal for Numerical Methods in Engineering}, Vol.~79,
  No.~11, 2009, pp. 1309--1331.
\newblock \doi{10.1002/nme.2579}.

\bibitem[{Winslow(1966)}]{winslow1966}
Winslow, A.~M., \enquote{Numerical solution of the quasilinear poisson equation
  in a nonuniform triangle mesh,} \emph{Journal of Computational Physics},
  Vol.~1, No.~2, 1966, pp. 149--172.
\newblock \doi{10.1016/0021-9991(66)90001-5}.

\bibitem[{Knupp(1999)}]{knupp1999}
Knupp, P., \enquote{Winslow Smoothing on Two-Dimensional Unstructured Meshes,}
  \emph{Eng. Comput-germany.}, Vol.~15, No.~3, 1999, pp. 263--268.
\newblock \doi{10.1007/s003660050021}.

\bibitem[{Brodersen(2002)}]{brodersen2002}
Brodersen, O., \enquote{Drag {Prediction} of {Engine}-{Airframe} {Interference}
  {Effects} {Using} {Unstructured} {Navier}-{Stokes} {Calculations},}
  \emph{Journal of Aircraft}, Vol.~39, No.~6, 2002, pp. 927--935.
\newblock \doi{10.2514/2.3037}.

\bibitem[{Berkelaar et~al.(2004)Berkelaar, Eikland, and Notebaert}]{lp_solve}
Berkelaar, M., Eikland, K., and Notebaert, P., \enquote{{lp\_solve} 5.5, Open
  source (Mixed-Integer) Linear Programming system,} Software, May 1 2004.
\newblock \urlprefix\url{http://lpsolve.sourceforge.net/5.5/}.

\end{thebibliography}

\end{document}